\documentclass[aps,pra,twocolumn,groupedaddress,floatfix,10pt]{revtex4-1}

\usepackage{amsmath}
\usepackage{bm}
\usepackage{graphics}
\usepackage{times}
\usepackage{graphicx}
\usepackage{xcolor}
\usepackage{array,multirow,graphicx}
\usepackage{multirow,hhline,graphicx,array}
\usepackage{rotating}
\usepackage{multirow}
\usepackage{booktabs}
\usepackage[markup=underlined]{changes}
\usepackage{soul}


\begin{document}

\title{Coherent control of multiphoton ionization of lithium atoms by a 
	 bichromatic laser field}
 
\author{S.~Me\v{z}inska$^{1}$}
\author{A.~Dorn$^{1}$}
\author{T.~Pfeifer$^{1}$}
\author{K.~Bartschat$^{2}$}

\affiliation{$^1$Max Planck Institute for Nuclear Physics, D-69117 Heidelberg, Germany}
\affiliation{$^2$Department of Physics and Astronomy, Drake University, Des Moines, IA 50311, USA}

\date{\today}

\begin{abstract}
We demonstrate a left-right asymmetry control of the photo\-electron angular distribution in multi\-photon ionization of Li atoms by a bichromatic laser field. By delaying the fundamental (780 nm) and its second harmonic relative to each other in steps of 130 atto\-seconds, we can vary the relative phase between the two laser fields with sub-wavelength accuracy and thereby steer the ejected electrons.
Good agreement is found between the measurements and calculations at the appropriate intensities of the two harmonics. 
\end{abstract}

\maketitle

\section{Introduction}\label{sec:Intro}

The interaction of atoms with intense laser fields is a key focus of fundamental physics~\cite{Agostini2004, Corkum2007, Krausz2009}, as complex dynamics arise out of seemingly ``simple'', well-understood atoms by their nonlinear interaction with a classical, but time-dependent, external electro\-magnetic field. Particular attention is paid to the control of these complex dynamics. A common approach to investigate these phenomena is the application of ionization schemes based on interference, which are realized using phase-locked two- or multi-color laser fields. Technically, a relatively straight\-forward way of achieving coherent control involves a two-color laser field at frequencies $\omega$ (fundamental) and $2\,\omega$ (second-harmonic). For example, it was shown that ionization with a linearly polarized two-color laser field \hbox{$\omega +2\,\omega$} can be employed to control the left-right asymmetry of the photo\-electron angular distribution (PAD) with respect to the laser polarization direction by changing the relative phase between the two harmonics.

The experimental demonstration of the left-right asymmetry control via such a scheme was reported already in 1992 by Elliott and coworkers~\cite{Yi-Yian1992}. They ionized atomic rubidium from its ground state via two-photon absorption at frequency $\omega$ and single-photon absorption at frequency $2\,\omega$ in the optical domain. They also suggested the potential of this particular ionization scheme for determining the phase difference between partial waves of opposite parities. The idea was further developed in the theory paper by Nakajima~\cite{Nakajima2000}, which in turn triggered further experimental work by Elliott \emph{et al.}, resulting in the measurement of the phase difference between the odd-parity $p$- and the even-parity $d$-wave~\cite{Wang2001}.

The ionization scheme based on simultaneous one- and two-photon ionization experienced a renaissance with interest in demonstrating the longitudinal coherence properties of free-electron laser (FEL) radiation at FERMI~\cite{Prince2016}. Several experiments were performed with neon starting from the 2$p$ sub\-shell. After first demonstrating the left-right asymmetry control of the PAD, more experiments followed with the extraction of several photo\-ionization parameters. In particular, by measuring the PADs as a function of the relative phase between the two harmonics and fitting them to a series of Legendre polynomials, the even- and odd-rank $\beta$ parameters could be determined. The $\beta$ parameters, in turn, enabled the extraction of the phase difference between the even- and odd-parity continuum wave functions~\cite{Fraia2019}. In addition, the FERMI group performed measurements of the emission-angle-dependent phase offset of the modulated photo\-electron signal, similar to the previous observations by Elliott and coworkers in the optical domain. By measuring these phases at three different photon energies, they determined the difference between the photo\-emission delays resulting from one-photon and two-photon ionization~\cite{You2020}. The above-mentioned experiments were performed in the multiphoton ionization regime with a Keldysh parameter~\cite{Keldysh1965} $\gamma_{\text{K}} \gg~1$. However, a left-right asymmetry control with an \hbox{$\omega+2\,\omega$} laser field was recently also demonstrated in the tunneling regime, where $\gamma_{\text{K}} \lesssim~1$~\cite{Arbo2015}.

In this paper, we present a left-right asymmetry control of the PAD of ground-state lithium atoms with respect to the relative phase of the components of a two-color laser field consisting of 780~nm (fundamental) and 390~nm (its second harmonic) radiation. However, in contrast to the above-mentioned experiments with rubidium and neon, our ionization scheme involves multi\-photon paths. 

This manuscript is organized as follows. We begin with the schematic setup of the study in Sect.~\ref{sec:Scheme}, followed by the description of the experimental setup in Sect.~\ref{sec:Experiment} and a brief summary of the theoretical method in Sect.~\ref{sec:Theory}.  The results are presented and discussed in  
Sec.~\ref{sec:Results}, before we finish with a summary and an
outlook in Sect.~\ref{sec:Summary}.

\begin{figure*}[t]
\centering
\includegraphics[width=0.245\linewidth]{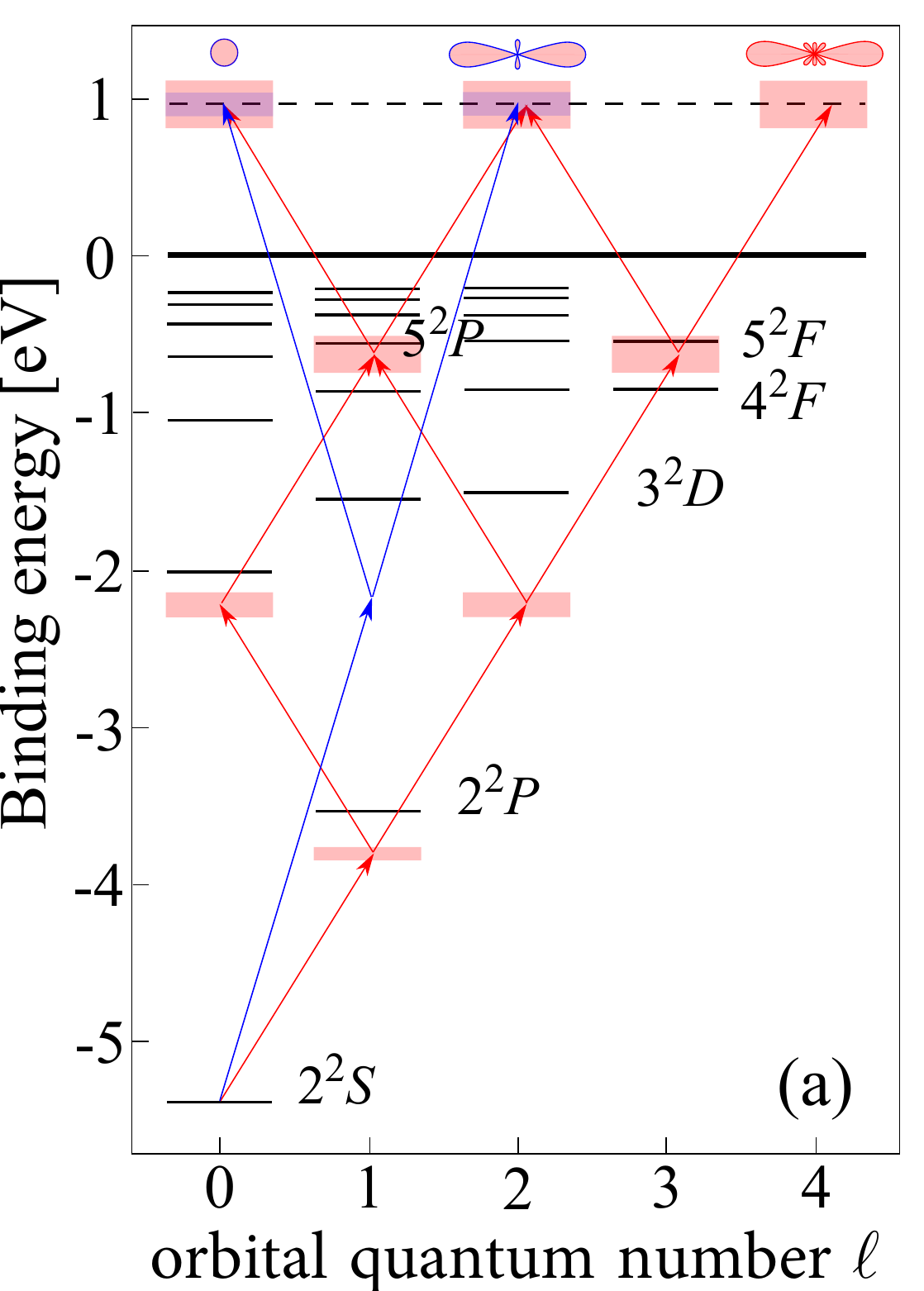}
\includegraphics[width=0.245\linewidth]{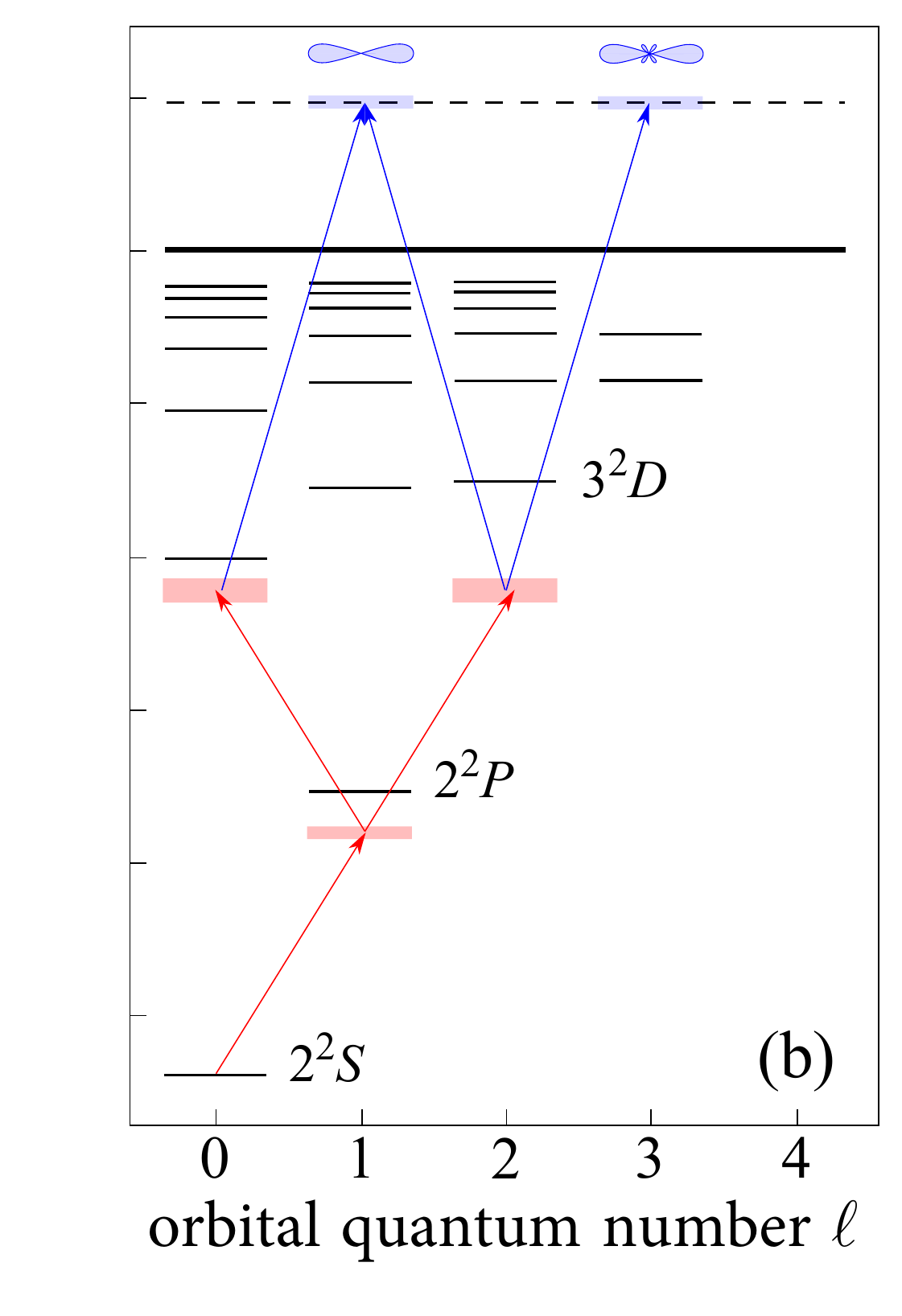}
\includegraphics[width=0.245\linewidth]{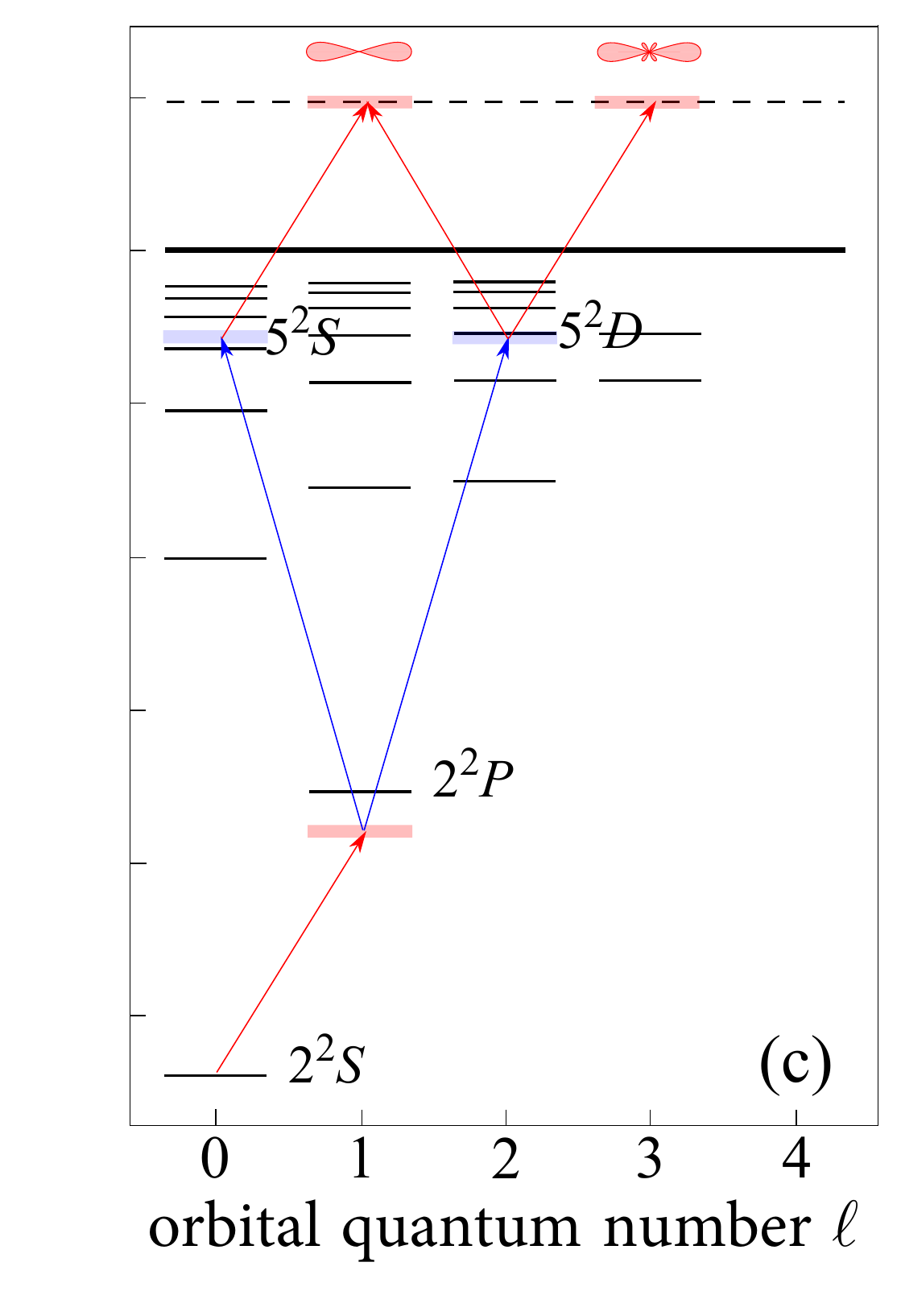}
\includegraphics[width=0.245\linewidth]{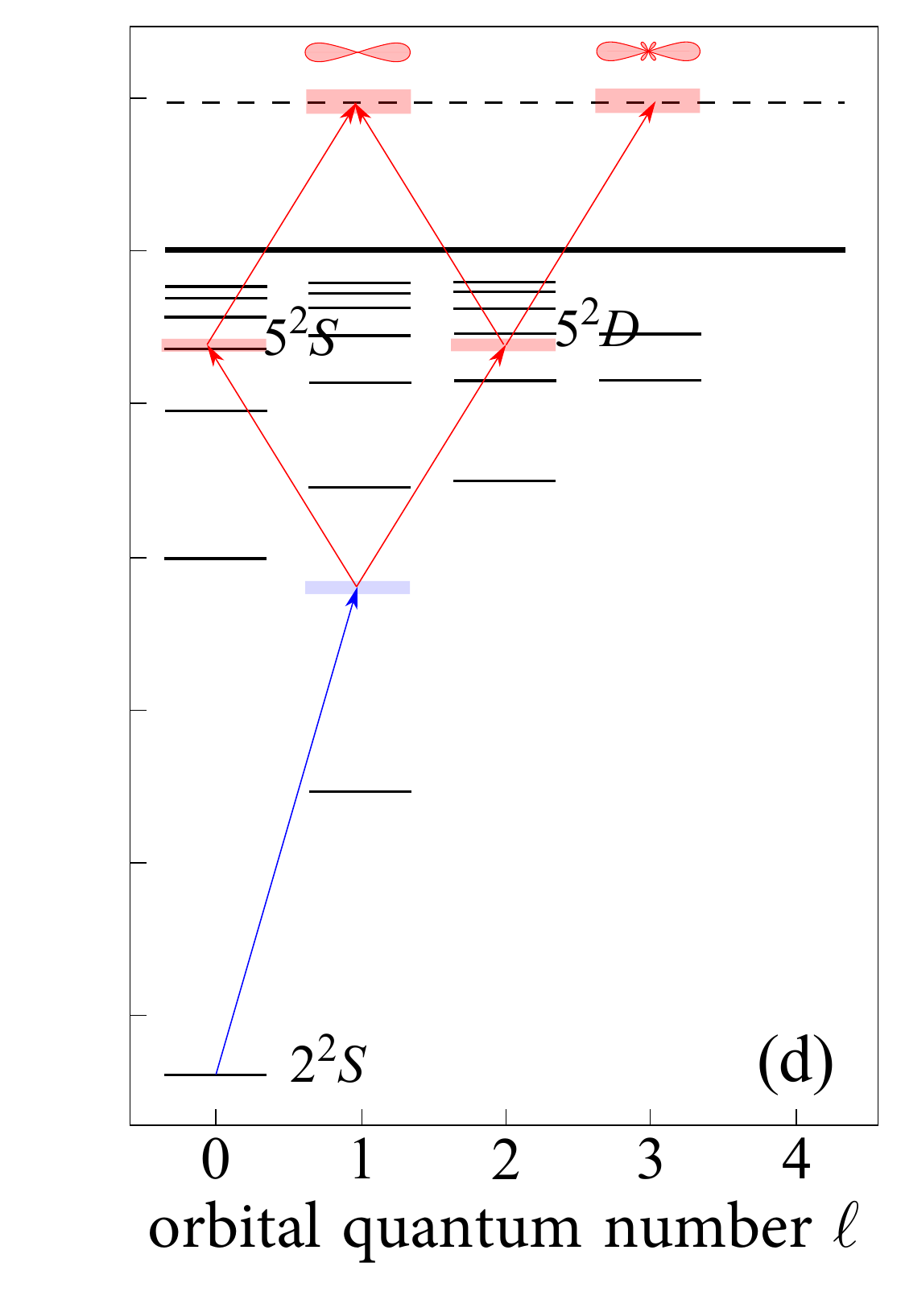}
\caption{Possible ionization paths. Panel (a) shows that the ejected-electron wavefunction has $s$-, $d$-, or $g$-symmetry (even parity) if either four red (780~nm) or two blue (390~nm) photons are absorbed.  The remaining panels are cases where two red and one blue photon are absorbed, thereby leading to continuum states of $p$- or $f$-symmetry (odd parity).}
\label{fig:paths}
\end{figure*}

\section{Scheme}\label{sec:Scheme}
In single-color multi\-photon ionization with linearly polarized light, the angular-momentum selection rules (\hbox{$\Delta \ell=\pm 1$}, \hbox{$\Delta m = 0$}) in the electric dipole approximation restrict the parity of the ejected electron to either even or odd, depending on the number of absorbed photons. Therefore, in single-color multi\-photon ionization, the continuum state has a well-defined parity. On the other hand, in two-color multi\-photon ionization with photon frequencies $\omega$ and $2\,\omega$, the parity of the wave function describing photo\-electrons with energy~$E$ may be undefined. In our ionization scheme of ground-state lithium Li$(2s)^2S$, the absorption of four fundamental photons generates even-parity continuum electrons with $s$-, $d$- and $g$-wave character (see Fig.1 (a)). Ionization by the second harmonic with two-photon absorption generates continuum $s$- and $d$-partial waves. On the other hand, two-color ionization paths corresponding to the absorption of one $2\,\omega$ photon and two fundamental $\omega$ photons produce odd-parity continuum electrons with $p$- and $f$-wave character 
(see \hbox{Fig.~\ref{fig:paths}$\,$(b,c,d})).

Each partial wave with an orbital quantum number $\ell$ is described by its corresponding Legendre polynomial $P_\ell(\theta)$, where $\theta$ is the emission angle of the photo\-electron with respect to the laser polarization direction. From the symmetry property 
\begin{equation}
    P_\ell(90^{\circ} + \theta) = (-1)^{\ell} P_\ell(90^{\circ} - \theta),
    \label{eq: Legendre_symmetry}
\end{equation}
it follows that the $P_\ell$ for even-parity partial waves are symmetric with respect to $\theta = 90^\circ$, but anti\-symmetric for odd-parity partial waves. 

In our experiment, we measured the PAD, which is proportional to the absolute angle-differential cross section defined by
\begin{equation}
    \frac{d\sigma}{d\theta} 
    = \bigg |\sum_{\ell=0}^{\ell_{\textrm{max}}} c_{\ell} \,\text{e}^{\textrm{i}\eta_{\ell}} P_{\ell}(\theta)\bigg |^2 = 
    \frac{\sigma_{\text{tot}}}{4\pi}
    \sum_{\lambda=0}^{2\, \ell_{\textrm{max}}}[1 + \beta_\lambda P_\lambda(\theta)].
    \label{eq: diff_cross_section}
\end{equation}
The $c_\ell$ are complex coefficients and $\eta_{\ell}$ is the phase\-shift of the partial wave with orbital angular momentum~$\ell$. In our case, it is comprised of the Coulomb phase plus a potential-scattering phase that accounts for deviations from a pure Coulomb potential with asymptotic charge~$Z_\textrm{asym} = 1$ due to the short-range screening of the 
full nuclear charge of Li ($Z = 3$) by the $(1s^2)$ core of Li$^+$. Furthermore, $\sigma_{\text{tot}}$ is the angle-integrated (total) cross section, and the $\beta_\lambda$ are asymmetry parameters. Therefore, a PAD can be expressed either as the modulus squared of the sum of Legendre polynomials of all partial waves involved or as a sum of $\beta$ parameters with only one Legendre polynomial per term.

It follows from Eqs.~(\ref{eq: Legendre_symmetry}) and~(\ref{eq: diff_cross_section}) that the PAD of a continuum state with a well-defined parity, as in single-color multi\-photon ionization, is symmetric with respect to the 90$^{\circ}$ emission angle. In this case, the PADs are fully described by only even-rank $\beta$ parameters. On the other hand, in the case of a continuum state with undefined parity, as in two-color multi\-photon ionization, a PAD can still be symmetric under very special circumstances, but it is generally asymmetric with respect to $\theta$ = 90$^\circ$. Asymmetric PADs are described by both even- and odd-rank $\beta$ parameters.

The degree of asymmetry depends on the relative phase $\Phi$ between the two laser fields at frequencies $\omega$ and $2\,\omega$ as well as the intensities of the two harmonics. In general, the relative intensities are chosen to obtain the best asymmetry contrast, while the control of the degree of asymmetry is realized by varying the relative phase $\Phi$ with sub-wavelength accuracy. The latter is a consequence of $\Phi$ being imprinted on the partial waves generated via the two-color paths (see Eq.~(2) in~\cite{Fraia2019}).

The degree of asymmetry can be quantified by means of the asymmetry parameter 
\begin{equation}
    A_\text{LR} \equiv \frac{I_\text{L} - I_\text{R}}{I_\text{L} + I_\text{R}}.
    \label{eq: asymmetry_parameter}
\end{equation}
Here $I_{\text{L}}$ is the integrated intensity of the PAD over emission angles \hbox{$0 \le \theta \le \pi/2$}, while $I_{\text{R}}$ is obtained by integrating over \hbox{$\pi/2 \le \theta\le \pi$}~\cite{Prince2016}.

\begin{figure*}[!t]
\centering
\includegraphics[width=0.50\linewidth]{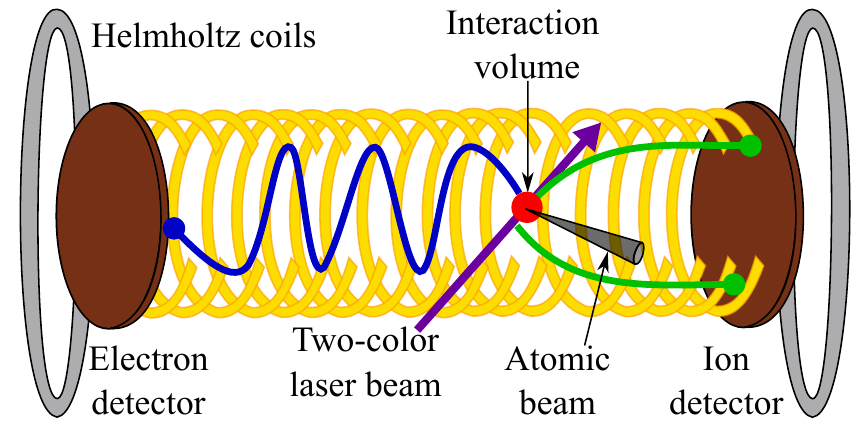}

\vspace{5.0truemm}

\includegraphics[width=0.90\linewidth]{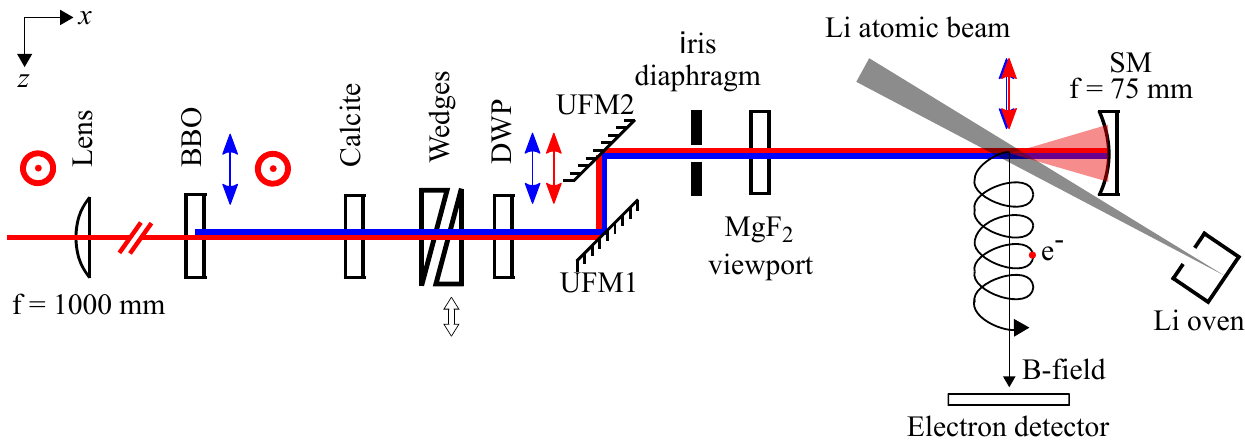}
\caption{Experimental setup. Top: ReMi apparatus. Bottom: Two-color optical setup. Lens: Plano-concave lens with \hbox{f = 1000 mm}; BBO: Beta Barium Borate crystal; Calcite: Calcite delay compensation plates; Wedges: A pair of dispersion wedges; DWP: Dual-wavelength waveplate; UFM1, UFM2: Dual-band ultrafast mirrors coated for 780~nm and 390~nm. MgF$_2$ viewport: vacuum viewport window of the ReMi. SM: Aluminum-coated concave spherical mirror with \hbox{f~= 75~mm}. Red and blue arrows: \hbox{p-polarization} of the fundamental and SH harmonic, respectively; Red circles: \hbox{s-polarization} of the fundamental radiation.}
\label{fig:experiment}
\end{figure*}

\section{Experimental Setup}\label{sec:Experiment}
We measured the PADs using a Reaction Microscope (ReMi), which is illustrated in the top portion of Fig.~\ref{fig:experiment}. The ReMi enables angle-resolved coincidence measurements of ions and electrons. In detail, the ions and electrons produced in the interaction volume, defined by the intersection between the atomic beam and the focal volume of the two-color laser beam, are extracted and guided to their respective detectors by applying an external homogeneous electric field, which is formed by a stack of electrodes. In addition, a homogeneous magnetic field is created by a pair of Helmholtz coils, which subjects electrons that are not ejected along the magnetic field direction to a cyclotron motion.

The bottom part of Fig.~\ref{fig:experiment} shows the optical setup for generating the two-color laser beam. We produce the second-harmonic (SH) radiation by a nonlinear Beta Barium Borate (BBO) crystal placed at the beam waist of a focused fundamental beam with a focal length of \hbox{f = 1000 mm}. The fundamental radiation is generated by a Ti:Sa femtosecond laser system operated at a 4~kHz repetition rate with a maximum pulse energy of 800~$\mu$J. The pulse duration (FWHM of intensity) is around 30~fs. Behind the BBO crystal, the co-propagating fundamental and SH beams are linearly polarized with orthogonal electric field vectors. We rotate the fundamental beam polarization along the SH beam using a dual-wavelength waveplate. Because of the different group velocities of the fundamental and SH pulses in the optical elements following SHG, they are temporally delayed relative to each other when arriving at the interaction volume. Specifically, the total delay introduced by the BBO crystal, the wedges, the dual-wavelength plate, the vacuum viewport window, and the air are compensated by calcite delay compensation plates. We control the relative phase between the harmonics with sub-wavelength accuracy by employing a pair of fused silica wedges, one of which is mounted on a translation stage. By changing its insertion depth, effectively a glass plate of variable thickness is formed, and the relative phase between the two harmonics can be varied in the desired way. The ReMi is equipped with a magneto-optical trap and an optical dipole trap to cool and store target atoms. In the present experiment, however, a simple collimated oven beam is used. 

\begin{figure*}[!t]
\includegraphics[width=1\linewidth]{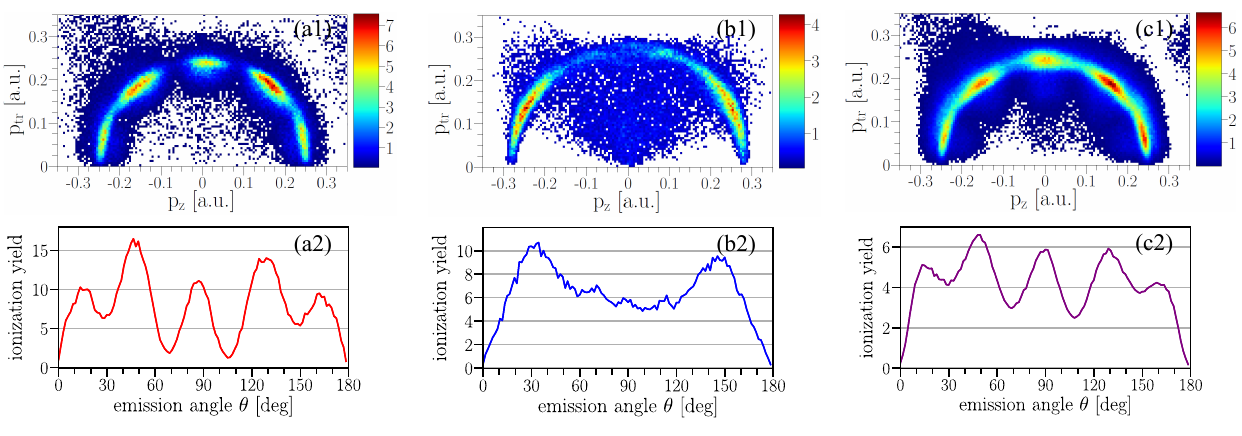}
\caption{Measured two-dimensional momentum distributions and angular distributions of ground-state lithium at 780~nm (a1, a2), 390~nm (b1, b2), and 780~nm / 390~nm (c1, c2).}
\label{im: 2Dmomentum}
\end{figure*}

Figures~\ref{im: 2Dmomentum} (a1-c1) display typical momentum spectra obtained by the ReMi, where the photo\-electron intensity is plotted as a function of the transversal momentum~p$_{\text{tr}}$ and the 
longitudinal momentum~p$_\text{z}$
of the photo\-electrons, respectively. In~(a1), the momentum spectrum shown results from ionization with only the fundamental radiation at 780~nm after four-photon absorption (Fig.~\ref{fig:paths}~(a)). In~(b1), it results from ionization with only the second-harmonic radiation at 390~nm after two-photon absorption (Fig.~\ref{fig:paths} (a)). Finally, two-color ionization with 780~nm + 390~nm radiation after three-photon absorption (Fig.~\ref{fig:paths} (b-d)) produces the pattern shown in~(c1).

The momentum components are defined with respect to the laser polarization direction. Specifically, p$_{\text{tr}}$ corresponds to the momentum perpendicular the laser polarization direction, whereas p$_\text{z}$ is parallel to it. We reconstruct the longitudinal momentum p$_\text{z}$ from the measured time-of-flight (TOF), and the transversal momentum from both the measured TOF and the $(x,y)$-coordinates of the photo\-electron detected by the time- and position-sensitive detector. From there, we can generate the PADs
studied in this work. For this, we integrated ionization yield over the total momentum range of p = 0.2 - 0.3 a.u. Figure~\ref{im: 2Dmomentum}~(a2) shows a typical PAD resulting from the ionization with 780 nm radiation, (b2) 390 nm, and (c2) 780 nm + 390 nm. The five maxima in (a2) and (c2) originate from the superposition of the $s$-, $d$- and $g$-partial waves with their characteristic PADs. The enhanced ionization yield at the minima in (c2) compared to (a1) arises from the $p$- and $f$- waves being generated in the two-color ionization. The PAD in (b2) shows two distinct maxima characteristic to the $d$-wave.

\section{Theory}\label{sec:Theory}
The experimental data are compared to predictions from {\it ab initio} calculations based on solving the  time-dependent Schr\"odinger equation (TDSE) in the single-active electron (SAE) approximation for the valence electron moving in a He-like $1s^2$ ionic core. 
The static Hartree potential~\cite{Albright1993,Schuricke2011} is calculated first and then supplemented by phenomenological terms to account for exchange with the core electrons as well as the core polarizability.
Details can be found in Ref.~\cite{Silva2021}. As shown earlier \cite{Silva2021b}, our model potential describes the 
atomic structure with an accuracy  of better than 1~meV for the entire Rydberg spectrum of neutral Li.

The two-color laser field was constructed by Fourier transforming our best estimate of the experimental frequency spectrum and accounting for the temporal composition of the pulses. Significant remaining uncertainties, however, concerned the absolute peak intensities and even the ratio between the fundamental and the second harmonic.  Consequently, several sets of calculations were performed to narrow down the likely parameter space.  Given the excellent agreement between previous calculations performed with the same computer code and experimental data measured under similar conditions~\cite{Silva2021,Silva2021b,Silva2021c}, we are confident in 
theory having sufficient predictive power in this respect. 

In theory, the PADs can be obtained either directly from the $\beta$ parameters or using the same procedure as applied to the experimental data, i.e., from the calculated longitudinal p$_\text{z}$ and transversal p$_{\text{tr}}$ momentum components of the photo\-electrons. Even though the latter procedure appears like a detour, we applied it in order to process the results in the same way as the experimental data, so that we could properly account for the experimental resolution.

\section{Results and Discussion}\label{sec:Results}
In this Section, we describe our results for the coherent control of the left-right asymmetry of the PADs, which is achieved by changing the relative phase $\Phi$ between the fundamental and SH components of the two-color pulses with sub-wavelength accuracy. 

\subsection{Left-right asymmetry}
\label{subsec: left_right_asymmetry}

\begin{figure}[!t]
\includegraphics[width=1\linewidth]{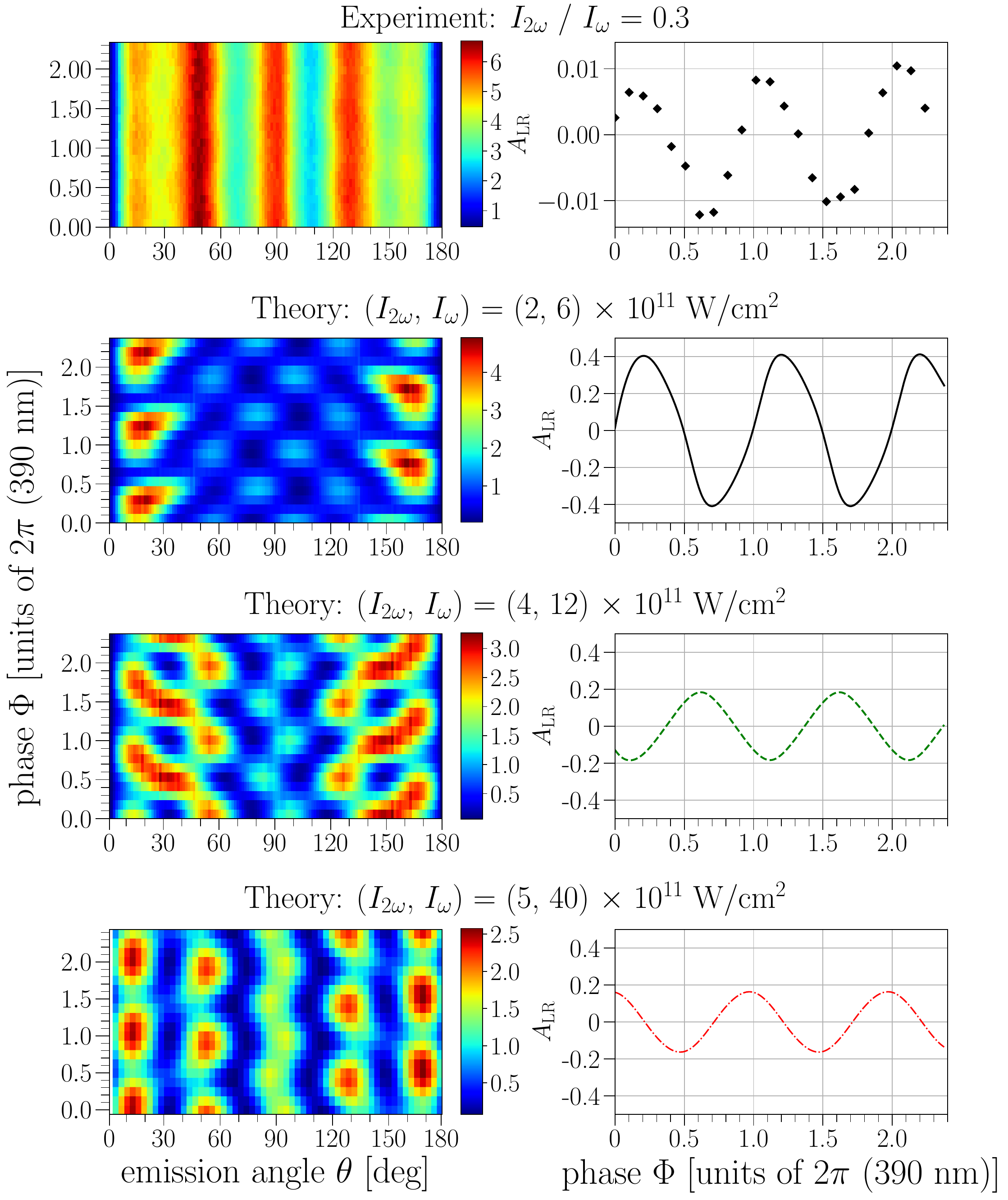}
\caption{Left column: raw measured (top) and calculated relative phase $\Phi$ traces of the PADs for selected peak intensities and intensity ratios. Right column: Corresponding angle-integrated asymmetry parameter $A_{\text{LR}}$ for each trace.}
\label{im: asymmetry}
\end{figure}

Figure~\ref{im: asymmetry} shows relative phase $\Phi$ traces of the measured and calculated PADs over two cycles of the blue light's optical period of 1.3~fs. The $\Phi$ step sizes in experiment and theory were 1/10 and 1/8 of that period, respectively. The calculated traces are presented at three different sets of the fundamental and SH pulse peak intensities. The measurements were performed at the $I_{2\omega}$/$I_{\omega}$ ratio of~$\approx 0.3$. 

Each trace is quantified in terms of the asymmetry parameter $A_{\text{LR}}$ introduced in Eq.~(\ref{eq: asymmetry_parameter}). We find that the $A_{\text{LR}}$ modulation phase of the measured trace agrees well with that of the calculated trace at $I_{2\omega} = 2 \times 10^{11}$ W/cm$^2$ and $I_{\omega} = 6 \times 10^{11}$ W/cm$^2$. However, the $A_{\text{LR}}$ modulation of the measured trace exhibits a significantly smaller modulation depth compared to theory. This is caused by the strong ionization with the fundamental radiation in our experiment due to imperfect matching of the two-color beam sizes and their profiles. We conclude this based on the PAD maxima, which are located at emission angles close to 45$^\circ$ and 135$^\circ$, which are characteristic for PADs generated only with the fundamental pulses (see Figure~\ref{im: 2Dmomentum} (a2)).  

Our measured PADs exhibit a small left-right asymmetry, which does not change with the relative phase $\Phi$ (see Fig.~\ref{im: 2Dmomentum}). This results from a reduced detection efficiency for the second electron if two electrons are produced from two atoms within the same laser pulse. This asymmetry leads to a small positive offset in the $A_{\text{LR}}$ modulation. We eliminated this experimental effect by setting the experimental mean asymmetry to zero.

The $A_{\text{LR}}$ modulation of the calculated PADs shows a strong dependence on the absolute intensities of the two harmonics. In recent theoretical studies on neon, such a strong dependence of the phase offset of $A_{\text{LR}}$ on the ionizing pulse intensities was found in the presence of resonant ionization transitions~\cite{Gryzlova2018}. In particular, a phase jump of $\pi$ is predicted close to a resonant intermediate state.

This scenario is also very likely in our ionization scheme of lithium due to so-called dynamic resonances, which are induced via AC Stark shifts of the ground state and the high-lying Rydberg states at different times and to various extents during the pulse depending on the laser peak intensities. The physical Rydberg states can then be shifted in or out of resonance with the virtual inter\-mediate state after three-photon absorption at frequency~$\omega$ or two-photon absorption at frequencies $\omega$ and $2\,\omega$ (see Fig.~\ref{fig:paths}). We found, indeed, that these dynamic resonances at various fundamental and SH pulse intensities used in the calculations favor the population of different Rydberg states. According to our estimates, the 4$^2P$, 4$^2D$, and 4$^2F$ states are populated significantly at \hbox{$I_{2\omega} = 2 \times 10^{11}$~W/cm$^2$} and \hbox{$I_{\omega} = 6~\times~10^{11}$~W/cm$^2$}, while the 4$^2S$, 4$^2P$, and 4$^2F$ states are populated at \hbox{$I_{2\omega} = 4~\times~10^{11}$~W/cm$^2$} and \hbox{$I_{\omega} = 12~\times~10^{11}$~W/cm$^2$}.  On the other hand,  no significant population of Rydberg states occurs at \hbox{$I_{2\omega} = 5~\times~10^{11}$~W/cm$^2$} and \hbox{$I_{\omega} = 40~\times~10^{11}$~W/cm$^2$}. The dependence of the Rydberg-state population on the intensity of the two-color laser field is thus the likely reason for the rapid changes in the phase offset of the $A_{\text{LR}}$ modulation. 

\begin{figure}[!b]
\includegraphics[width = 0.85\linewidth]{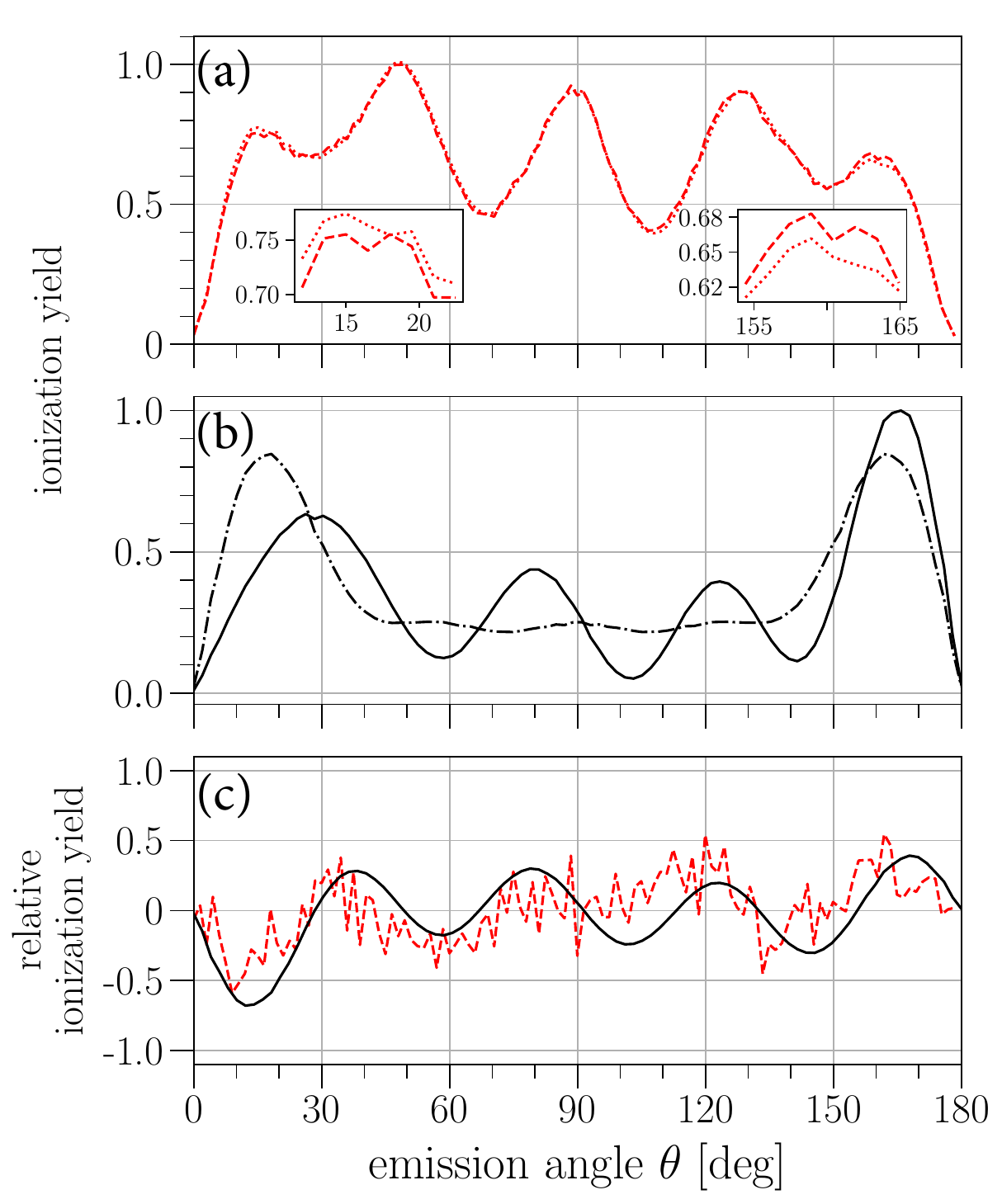}
\caption{(a) Raw measured PAD (dashed curve) and average of measured PADs (dotted curve). (b) Raw calculated  PAD (solid curve) and average of calculated PADs (dash-dotted curve). (c) Processed measured (dashed curve) and calculated (solid curve) PADs of the raw data in (a) and (b) according to the steps described in the main text. In (a) and (b), the ionization yields of the PADs are normalized with respect to the peak ionization yield of the raw data. The PADs shown correspond to $\Phi$ = 0.8.}
\label{im: background_subtraction}
\end{figure}

\begin{figure*}[!t]
\centering
\includegraphics[width=0.4\linewidth]{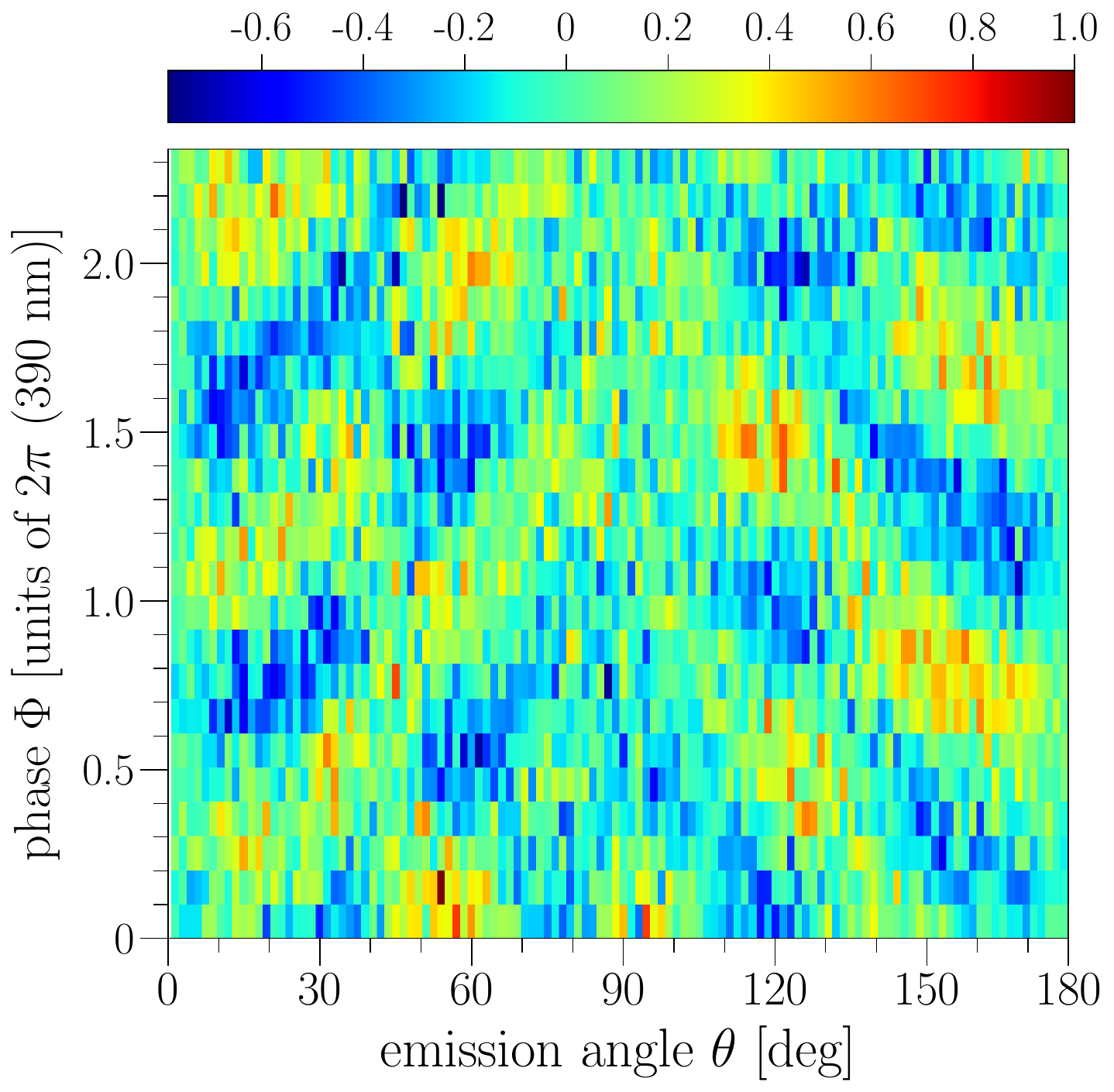}
\includegraphics[width=0.4\linewidth]{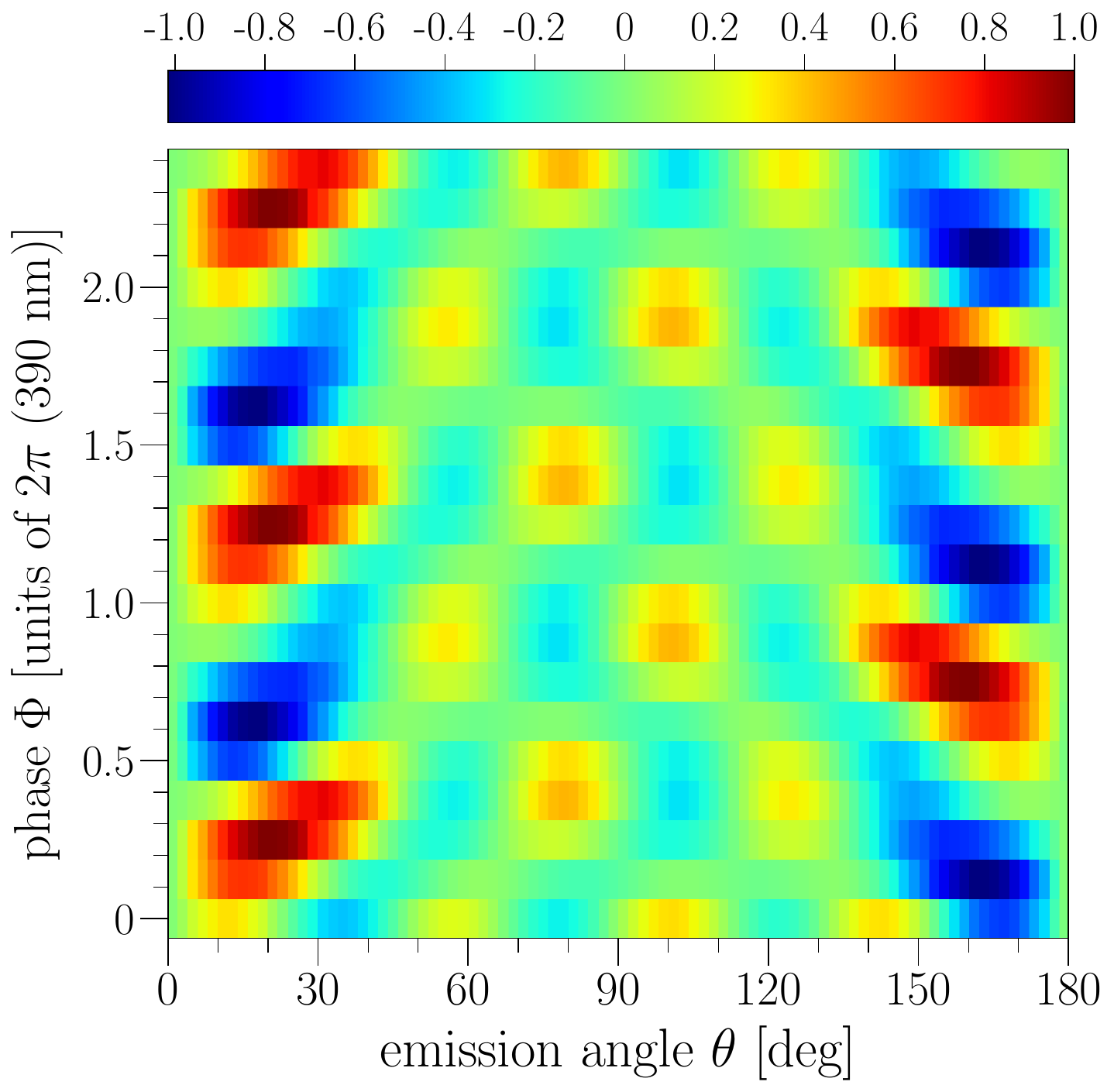}
\caption{Measured (left) and calculated (right) relative phase $\Phi$ traces of the PADs with the background subtracted as described in the main text.}
\label{im:2D_pad_theo_exp}
\end{figure*}

Since the measured PADs contain a large background, they cannot be directly compared with the calculated ones. Therefore, to make this possible, we processed the measured and calculated PADs at \hbox{$I_{2\omega} = 2 \times 10^{11}$~W/cm$^2$} and \hbox{$I_{\omega} = 6 \times 10^{11}$~W/cm$^2$} (see Fig.~\ref{im: background_subtraction}(a,b)) in the following way: We first calculated the average PAD over the entire relative phase range of the trace and then subtracted it from the individual PADs measured at each delay step. Finally, we normalized the results by setting the peak difference between the raw and the averaged data to unity. For theory, the locations of the predicted innermost minima and maxima do not change after subtracting the average PAD from the original PAD. On the other hand, the maxima and minima  of the measured PAD are shifted and now agree well with those of the raw and processed calculated PADs, as seen in Fig.~\ref{im: background_subtraction}(c).

Figure~\ref{im:2D_pad_theo_exp} displays the measured and calculated relative phase $\Phi$ traces with the subtracted background, which show good agreement. No modulation occurs at $90^\circ$, because all odd-rank Legendre polynomials vanish at that angle. 
Similarly, there is no modulation near 70$^\circ$/110$^\circ$, where the \hbox{$g$-wave} has a node. This indicates that interference with this wave is important for producing the modulation in the neighboring angular range. 
Since the odd-parity $f$-wave has a node at 40$^\circ$, the modulation in the data in this angular region is mainly due to the odd-parity $p$-wave contributing. At 55$^\circ$ the $d$-wave has a node.  In this case, the strong modulation is due to the contribution from the even-parity waves being reduced.  Finally,  the maxima near 60$^\circ$/120$^\circ$ agree with the maximum of the $f$-wave.

\subsection{Ionization yield modulation at various emission angles}

\begin{figure}[!t]
\includegraphics[width = 1\linewidth, height = 0.97\linewidth]{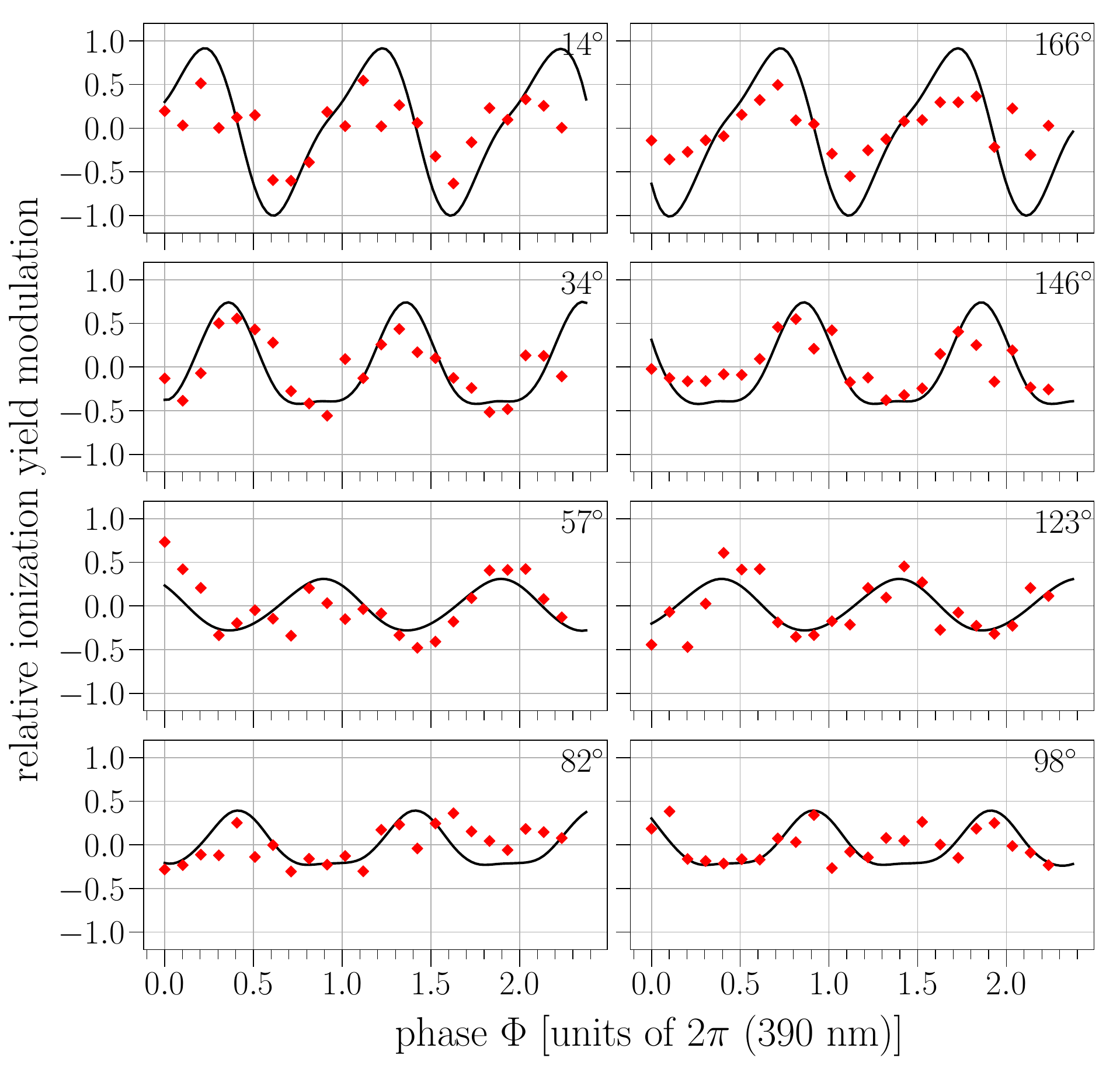}
\caption{Relative ionization yield modulation at various emission angles $\theta$ as a function of the relative phase $\Phi$. Red diamonds: experimental data points. Black lines: theory.}
\label{im: pad_modulation}
\end{figure}

Figure~\ref{im: pad_modulation} shows relative ionization yield modulations at various emission angles~$\theta$ and their counter\-parts \hbox{$180^\circ - \theta$}. These data correspond to those presented in Fig.~\ref{im:2D_pad_theo_exp}.  The modulated signals not only depend on the relative phase~$\Phi$ between the two colors, but also on the emission angle. Each of the modulated signals exhibits a phase offset, which varies with the emission angle. Ideally, the curves for $\theta$ and \hbox{$180^\circ - \theta$} should be shifted by~$\pi$ when the relative phase~$\Phi$ between the two colors is varied.  This is, of course, the case for the theoretical predictions, while some deviations are seen in the experimental data.  These deviations provide an indication about statistical and systematic uncertainties in the measurements. We also note that some binning ($\pm 2^\circ$) was applied to both the experimental data and the theoretical predictions in order to obtain a sufficiently large count number. 

\label{subsec: ionization_yield_modulation}
\begin{figure}[!t]
\includegraphics[width = 1\linewidth]{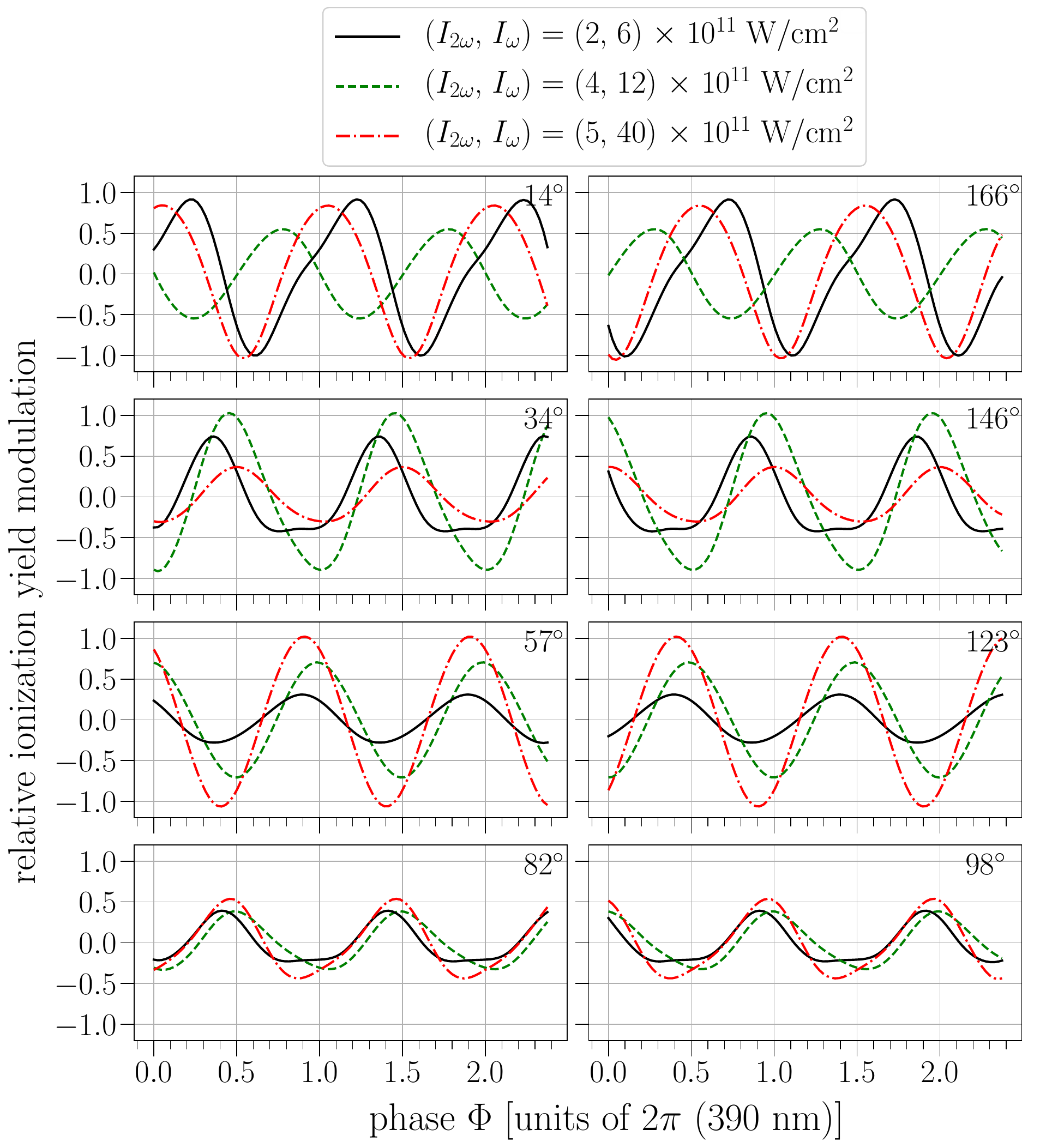}
\caption{Calculated dependence of the ionization yield modulation on the relative phase $\Phi$ at various emission angles $\theta$ for different absolute intensities and intensity ratios of the two harmonics. The data shown correspond to those in Fig.~\ref{im: asymmetry} and are processed in the same way as in Fig.~\ref{im:2D_pad_theo_exp}.}
\label{im: pad_modulation_theory}
\end{figure}

The above characteristics were also observed and discussed in earlier experiments for other target systems, where the left-right asymmetry in the PADs was controlled via interference between one- and two-photon ionization pathways~\cite{Yi-Yian1992,Fraia2019,You2020}. Therefore, we only briefly refer to the main ideas reported in the earlier works. While the $\pi$-shifted oscillations result from the symmetry properties of the two-color laser field~\cite{You2020}, the emission-angle-dependent phase offsets are directly related to the differences of the phase shifts $\eta_\ell$ of the partial waves that determine the emission at specific angles according to Eq.~(\ref{eq: diff_cross_section})~\cite{Yi-Yian1992, Fraia2019}. 

In our ionization scheme of lithium (see Fig.~\ref{fig:paths}), several intermediate resonant transitions to high-lying Rydberg states are possible. Recent experimental and theoretical studies on the variation of the phase offsets of the modulated ionization signals with the emission angle $\theta$ showed that in the case of the non\-resonant multiphoton ionization, the phase offset changes smoothly with the emission angle, whereas it exhibits rapid changes in the presence of intermediate resonances~\cite{You2020}. In the resonant case, theory also predicts a strong dependence of the phase offset on the absolute intensities of the two harmonics. This is indeed observed in the predictions shown in Fig.~\ref{im: pad_modulation_theory}, although it is yet to be 
confirmed experimentally. In particular, the calculations reveal that the strongest variation of the phase offset with intensity occurs along the polarization direction, whereas only a small variation is observed at 90$^\circ$. 

\section{Conclusions and Outlook} \label{sec:Summary}
We have demonstrated a left-right asymmetry control of the PADs of ground-state lithium using a bichromatic laser field consisting of 780~nm and 390~nm radiation. Except for the low asymmetry contrast in our measurements due to the large background caused by strong ionization at the fundamental wavelength, we found good agreement with theoretical predictions if appropriate absolute intensities and intensity ratios of the two harmonics were selected. Our calculations predict a strong intensity dependence of the emission-angle-dependent phase offsets of the ionization yield modulation and the asymmetry parameter. We currently attribute this dependence to dynamical resonances with intermediate Rydberg states. However, further experimental investigations would be required to unambiguously confirm this conclusion. 

\vspace{-0.5truecm}

\begin{acknowledgments}
The theoretical part of this work (K.B.) was supported by the NSF through grant \hbox{No.~PHY-2110023}.
Some of the calculations were performed on Frontera at the Texas Advanced Computing Center in Austin (TX) through the Pathways allocation PHY-20028, and on Expanse at the 
San Diego Super\-computing Center in San Diego (CA) through the ACCESS allocation MXCA08X034.
\end{acknowledgments}

\bibliographystyle{apsrev4-1}

\end{document}